\documentclass[doublecol]{epl2} 
\usepackage{graphicx,color,amsmath}

\newcommand{\phI}{Z$_{\rm a}$}
\newcommand{\phII}{Z$_{\rm d}$}
\newcommand{\phIII}{U$_{\rm ad}$}
\newcommand{\phIV}{U$_{\rm dd}$}

\newcommand{\figxg}{
\begin{figure}[t]
	\centering
	\includegraphics[width=2.5in]{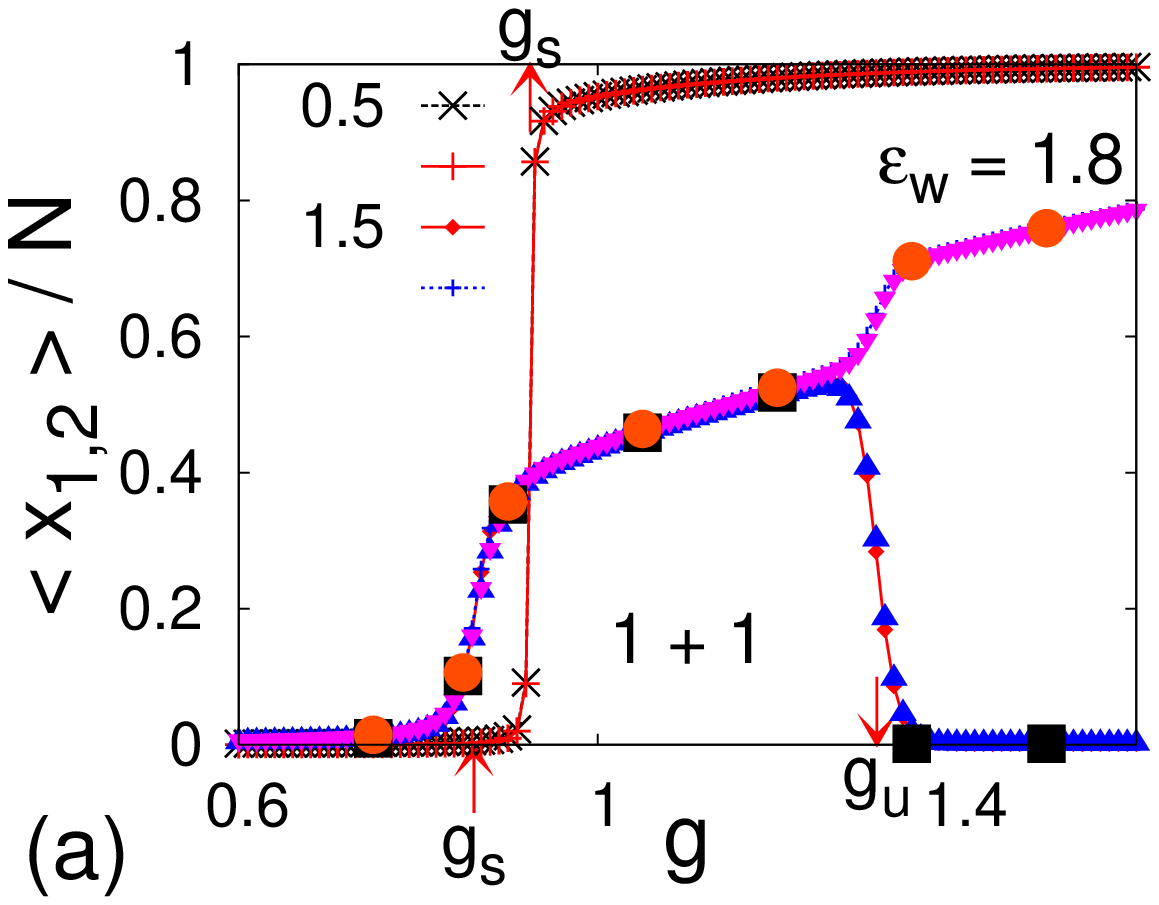}
	\includegraphics[width=2.5in]{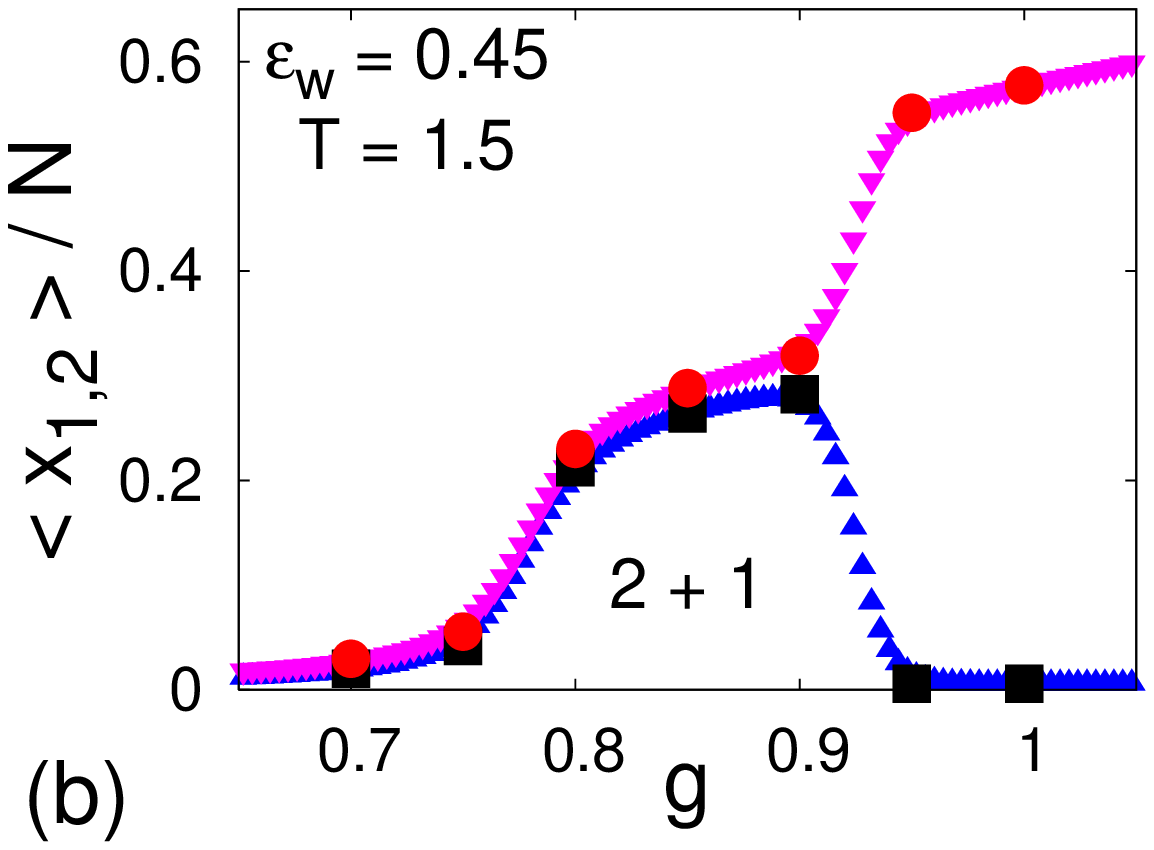}
	\includegraphics[width=2.5in]{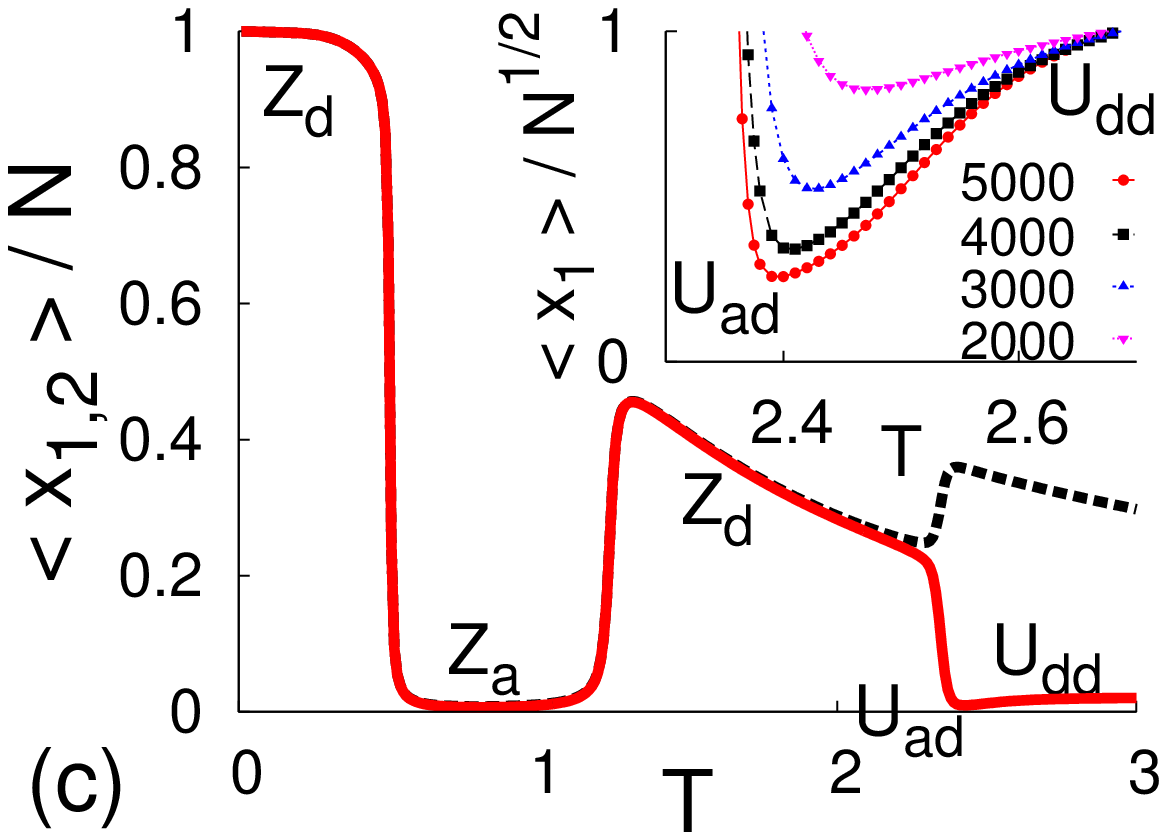}
 
	\caption{ The $\langle x_{i} \rangle /N$ vs $g$ ($i=1,2$) isotherms
	for $N=1000$ (a) for the $1+1$ dimensional case at temperatures,
	$T=0.5$ and $1.5$ for $\epsilon_w = 1.8$. (b) for the $2+1$
	dimensional case. In both (a) and (b), the big squares (for $x_1$)
	and circles (for $x_2$) show the averages obtained by Monte Carlo
	simulations and the upper and the lower triangles are the estimates
	given by the multiple histogram technique at various $g$.  (c)
	$\langle x_{1,2} \rangle /N$ vs $T$ for the DNA of length $N=3000$
	at $g = 0.925$.  The solid (dashed) line represents the free
	(pulled) strand.  The inset shows $\langle x_{1} \rangle /\sqrt{N}$
	vs $T$ for various chain lengths.  }\label{fig:xg}

\end{figure} 
}

\newcommand{\figtu}{
\begin{figure}[b] 
	\centering
	\includegraphics[width=2.5in]{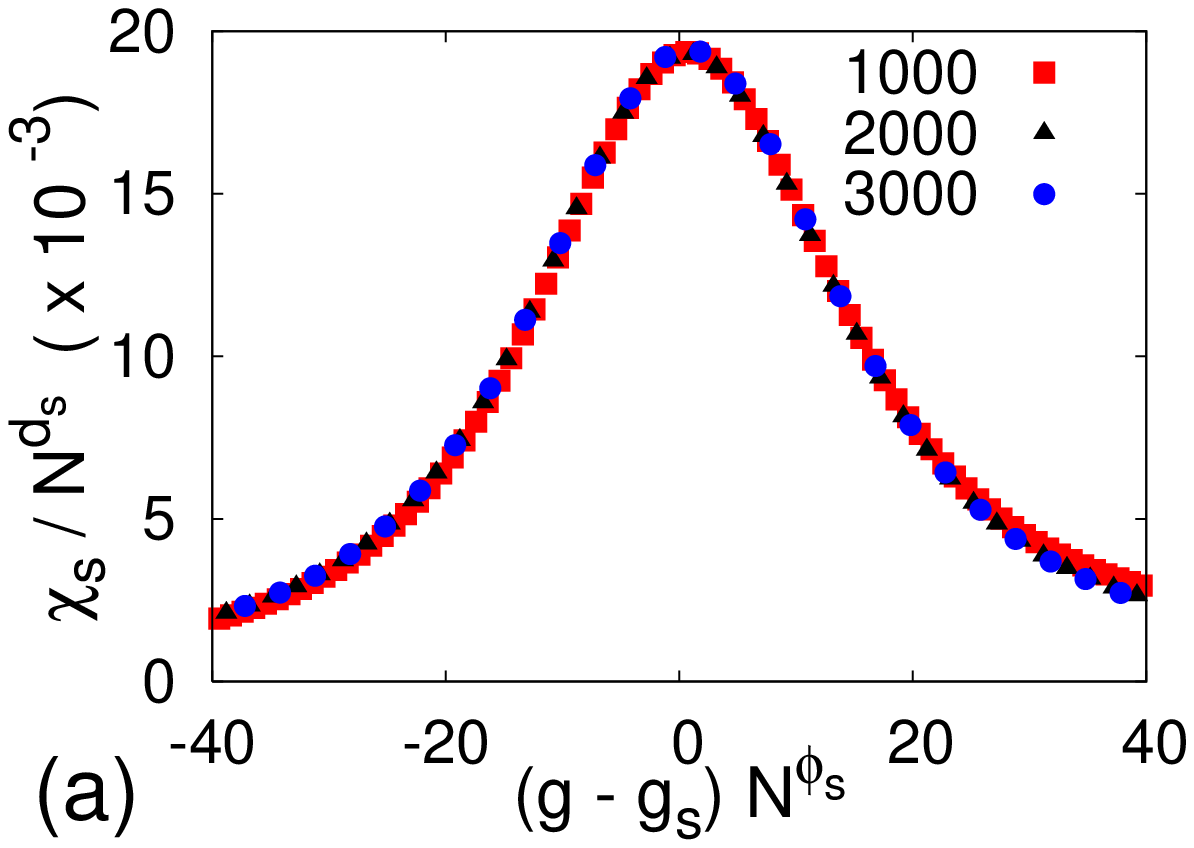}
	\includegraphics[width=2.5in]{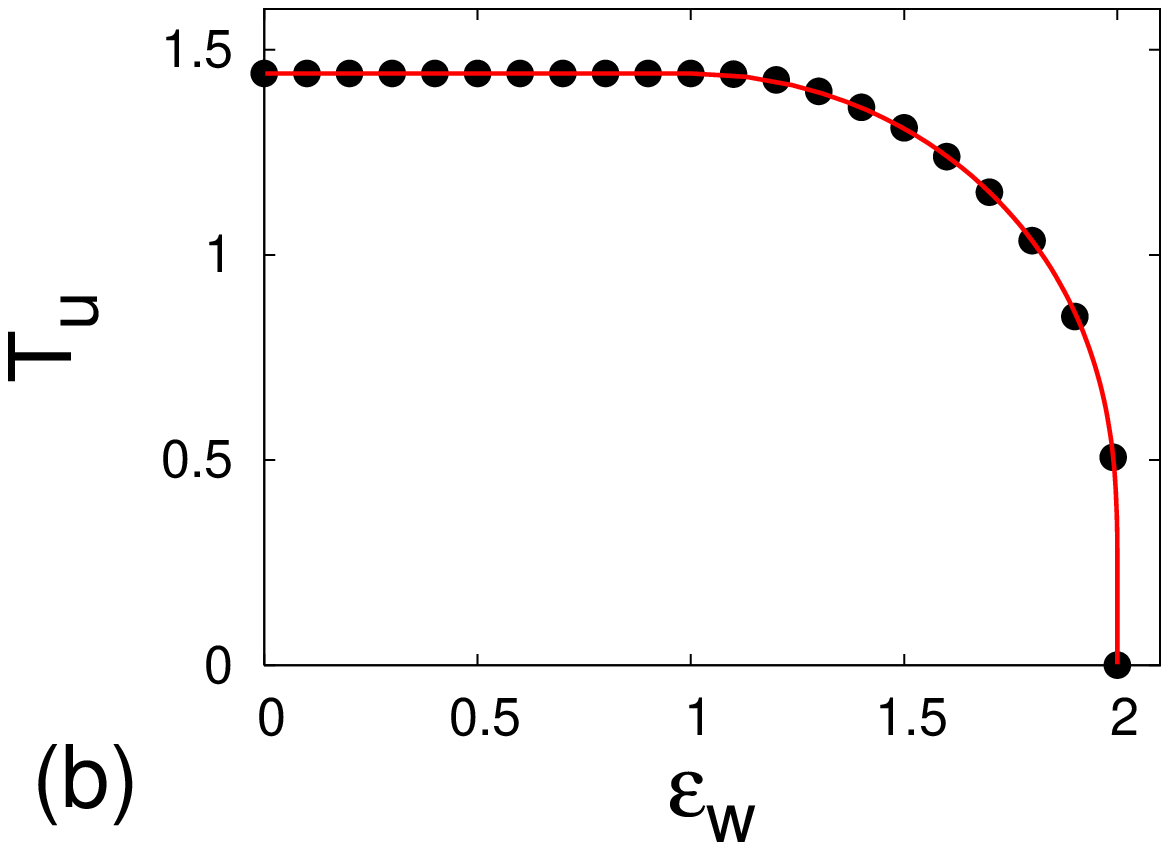}

	\caption{ (a) The nature of data collapse of the extensibility at
	$T=1.5$ for $N=1000$, $2000$ and $3000$ at transition Sz. (b) $T_u$
	vs $\epsilon_w$ curve from data collapse (points) and the exact
	curve (solid line) from Eq.~(\ref{eq:5}).  }\label{fig:tu}

\end{figure} 
}
        
\newcommand{\figmodel}{ 
\begin{figure}[b] 
	\centering
	\includegraphics[height=1.1in]{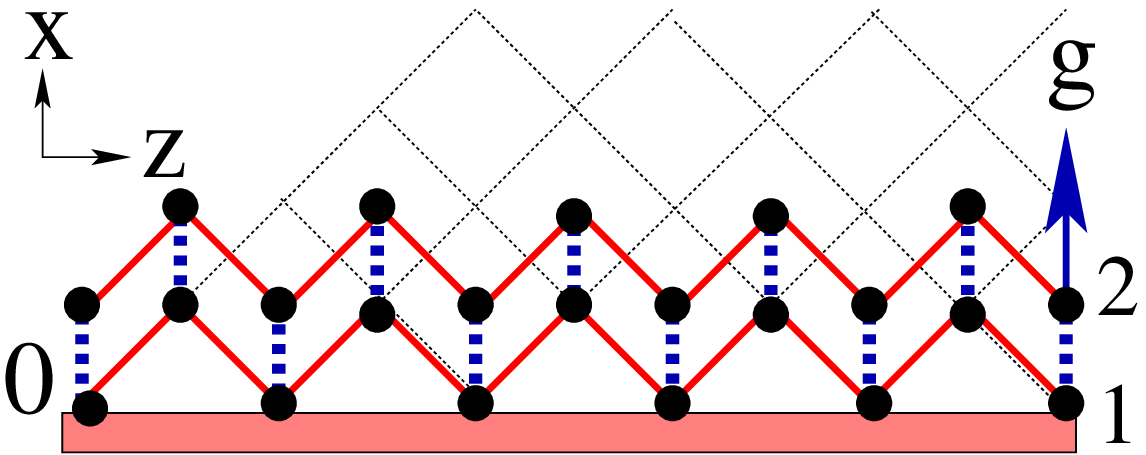} 

	\caption{ Schematic diagram of DNA adsorbed on the surface (shaded
	region). One end of the DNA is always kept anchored on the surface
	at the origin. The free strand (denoted by 1) can gain energy
	$-\epsilon_w$ for every contact with the surface (i.e.  $x_1 = 0$).
	An external force ${\mathrm g}$ (shown by arrow) is applied at the
	free end of the pulled strand (denoted by 2 and shifted by a unit
	distance to make it visible). The bold dotted lines denote the base
	pairing (energy $-\epsilon_b$) between the two strands of the DNA.
	For all figures, we take $k_B=1$ and $\epsilon_b = 1$.
	}\label{fig:model}

\end{figure} }

\newcommand{\figphdia}{
\begin{figure}[btp]
	\centering
	\includegraphics[width=2.5in,clip]{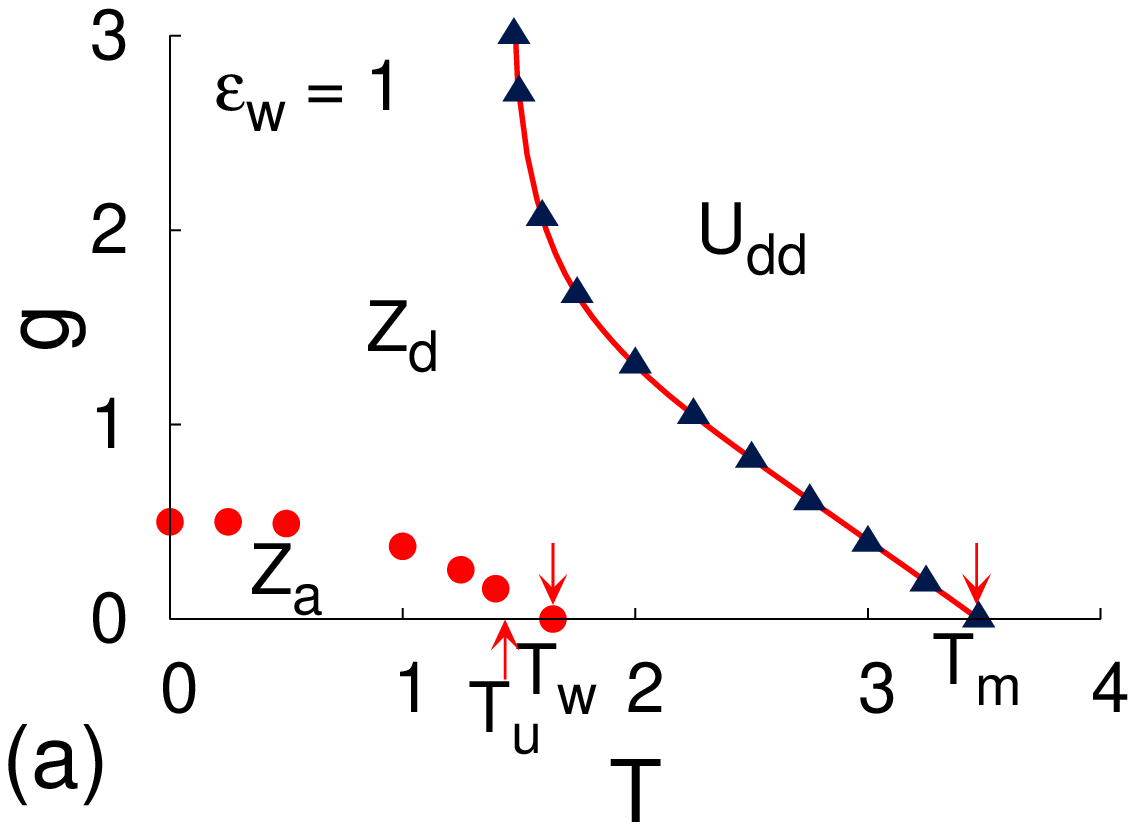}\\[18pt]
	\includegraphics[width=2.5in,clip]{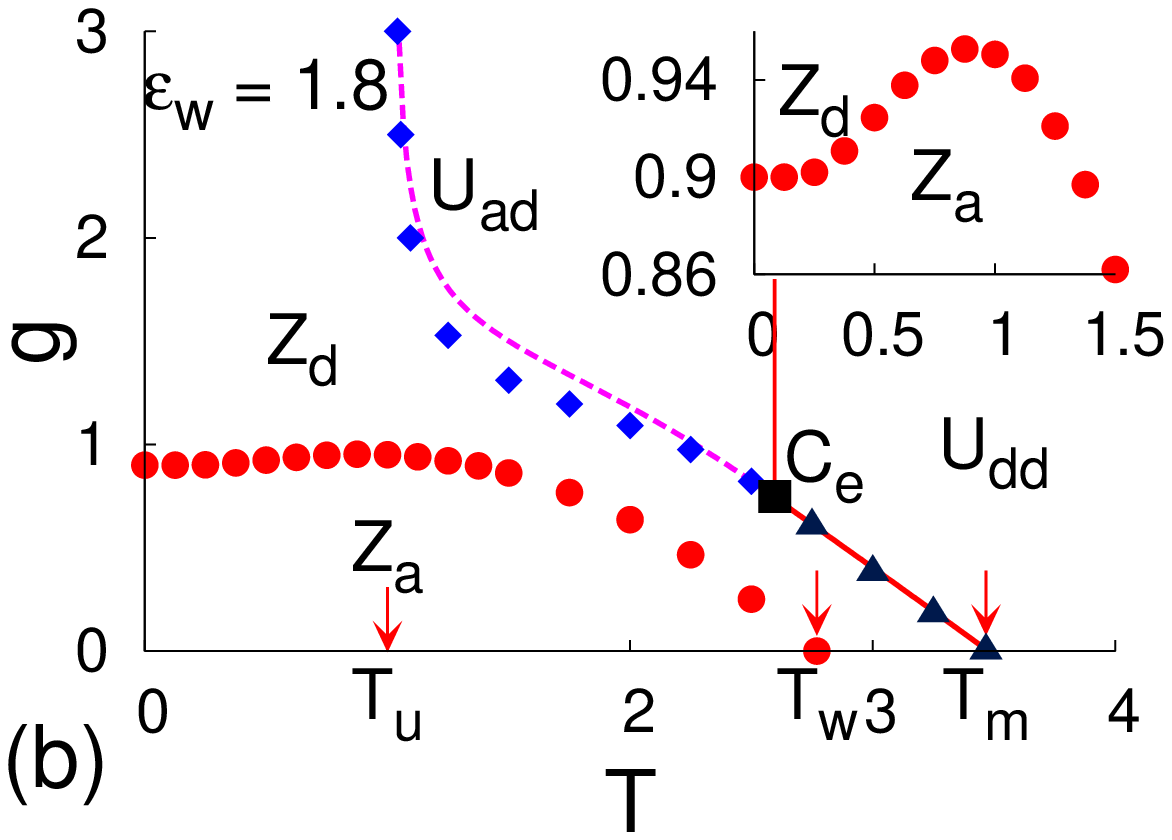}

	\caption{ $g$ vs $T$ phase diagrams (a) for $\epsilon_w = 1$ and (b)
	for $\epsilon_w = 1.8$. C$_{\rm e}$ represents the critical end
	point.  The  reentrance on the phase boundary separating phases \phI
	~and \phII ~is shown in the inset.  The points are from the transfer
	matrix and lines are Eqs.~(\ref{eq:3}) and (\ref{eq:4}).
	}\label{fig:phdia}

\end{figure} 
}

\newcommand{\figcecurv}{
\begin{figure}[btp]
	\centering
	\includegraphics[width=2.5in,clip]{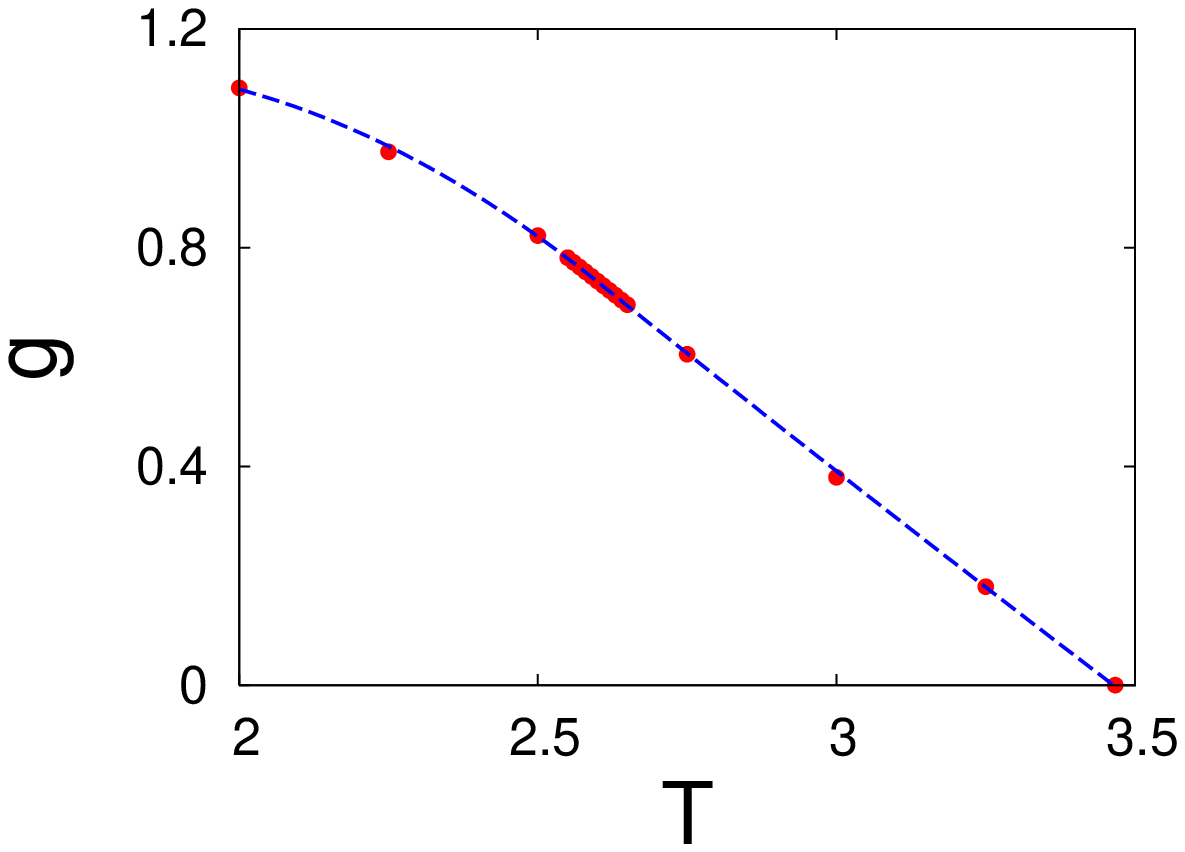}
	
	\caption{ Phase diagram in the neighbourhood of C$_{\rm e}$ to show
	the difference in curvature of the two phase boundaries.  The lines
	are the fits of Eq. (\ref{eq:2}) and the points are from the
	transfer matrix calculations.  }\label{fig:cecurv}

\end{figure} 
}

\title{Manipulating a single adsorbed DNA for a critical endpoint}
\author{Rajeev Kapri\inst{1}\thanks{Email:\email{rkapri@theory.tifr.res.in}} 
\and Somendra M. Bhattacharjee\inst{2}\thanks{Email:\email{somen@iopb.res.in}} }

\shortauthor{Rajeev Kapri and Somendra M. Bhattacharjee}

\institute{                    
  \inst{1}  Department of Theoretical Physics, Tata Institute of
  Fundamental Research, Mumbai 400 005. \\
  \inst{2}  Institute of Physics, Bhubaneswar 751 005 India. 
}

\abstract{ We show the existence of a critical endpoint in the phase
diagram of unzipping of an adsorbed double-stranded (ds) polymer like
DNA. The competition of base pairing, adsorption and stretching by an
external force leads to the critical end point.  From exact results, the
location of the critical end point is determined and its classical
nature established.  }

\begin{document}

\maketitle

A critical end point (CEP) occurs when a continuous transition line
terminates on a first order transition line.  It's specialty: even
though it is surrounded by three phases, still there is no three phase
coexistence, but, instead, a scale-free critical phase coexists with a
noncritical phase~\cite{helena,Landau}. For comparison, a triple point,
also surrounded by three phases, would show three phase coexistence. A
CEP is expected to occur in various mixtures and ferroelectrics and in
vortex lattice~\cite{helena,Landau,safar}.  We show here that a
different way of locating and studying a CEP is through single molecular
manipulations of an adsorbed DNA.

The melting of DNA is known to be a crucial step in many biological
processes~\cite{watson}.  The double stranded DNA(dsDNA) is a bound
state of two polymers or strands held together by hydrogen bonds of base
pairs.  The phenomenon of cooperative breaking of the base-pairings
thermally or otherwise is melting. The recognition of force as a
thermodynamic variable for this process has helped in completing the
phase transition picture of dsDNA~\cite{somen:unzip}.  Even though the
nature of the thermal denaturation of DNA remains a puzzle, the force
induced unzipping transition is theoretically
well-settled~\cite{somen:unzip,sebastian,maren,KapriSMB,scaling,Lubensky,Chen,kierfeld,KapriSMB:JPCM,Lam,GiriKumarPRE,Kapri:prl07,Kapri:physica,KumarGiriPRL}.
In addition, the force induced unzipping transition has emerged as a
possible scenario for opening of DNA~\cite{smb04} with the replication
Y-fork as the junction of two phase coexistence, the zipped and the
unzipped DNA. A thermodynamic description of DNA would entail two
conjugate ensembles of fixed force and fixed distance.  These two
ensembles are important in both theoretical and experimental
situations~\cite{KapriSMB,Keller,Roulet,Danilowicz}.

The Melting and unzipping of DNA are generally considered in the free
environment of bulk solutions, but often the presence of interacting
surfaces cannot be ignored.  {\it In vivo}, during replication, DNA gets
attached to the membrane but otherwise it remains away (``desorbed'')
from the membrane.  The protein-induced membrane-DNA attachment is used
in the replication process and cell division~\cite{firshein}.  In gene
therapy, targeted delivery is achieved by taking advantage of
adsorption-desorption of DNA on cationic
liposomes\cite{liposome1,liposome2}.  That metallic (e.g., gold),
semiconducting (e.g., silicon) or insulating (e.g., mica) surfaces can
also adsorb DNA has opened up the possibility of biosensors for fast and
precise detection of DNA in samples like hair, blood etc.  In all these
cases, the surface-DNA interaction depends (and hence tunable) on the
nature of the surface, fluctuation of the surface as for fluid
membranes, ionic concentration of the environment, nature of
hydrophobicity and van der Waal interactions.  A well-studied system in
recent years is DNA on gold where the DNA can be attached to the surface
with a thiol group and a small linker~\cite{sensor,eckel,mica}. The
model setup we are considering is similar to the case of DNA on a gold
plate and is shown in Fig.~\ref{fig:model}.  A force pulls only one of
the two strands of a DNA which can adsorb on an attractive surface. The
DNA as usual has the base pairing energy.  We here treat the surface-DNA
interaction as an additional parameter in the problem. It transpires
that for certain ranges of the attraction with surface there would be a
competition between adsorption, unzipping, and melting.  These three
processes can lead to a CEP.  The essential feature that holds the key
for the CEP is that the force-induced unzipping is strictly first order
but the adsorption-desorption transition of a polymer from a surface is
continuous~\cite{kapri:pre05,orlandini,mks05}.

\figmodel

Previous studies of DNA unzipping showed that the lattice model
preserves, even in two dimensions, the basic results of DNA unzipping
including the first order nature of the phase transition and the
existence of a reentrant region~\cite{maren}. The generic arguments of
these studies also showed that the choice of the lattice is not crucial.
For the problem at hand, we have also done Monte Carlo simulations in
$2+1$ dimensions (on a cubic lattice) and find that the force-distance
isotherms are qualitatively similar to the isotherms obtained in $1+1$
dimensions (see below).  Consequently we focus on the  results obtained
from analysis of exact results in 1+1 dimensions.

We model the DNA by two directed self avoiding walks on a $D=1+1$
dimensional square lattice. See Fig.~\ref{fig:model}. The walks, labeled
1 and 2, starting from the origin, are directed along the diagonal of
the square ($z$ direction). The walks are not allowed to {\it cross each
other} but whenever they meet (i.e.  $x_1(z) = x_2(z)$) there is a gain
in energy $-\epsilon_b (\epsilon_b > 0)$ for every contact. This is the
base pairing. At the diagonal ($x=0$) there is an impenetrable
attractive surface, with an energy $-\epsilon_w (\epsilon_w >0)$, which
favours the adsorption of the DNA.  In $1+1$ dimensions, the surface is
a line passing through the diagonal of a square lattice, and only one of
the strands can get adsorbed on it (i.e. $x_1 = 0$), since the two
strands of the DNA cannot cross each other. One end of the DNA is always
kept anchored at the origin. We apply an external force $g$, along the
transverse direction ($x$-direction) on the free end of one of the
strands of the DNA.  The other strand is left free.  Henceforth, the
strand which is left free is called the ``free strand" and the strand,
on which the external force acts is called the ``pulled strand".  The
endpoint positions $x_i(N)$ is to be shortened to $x_i$.  In $2+1$
dimensions, the surface is a plane passing through the diagonal of a
cubic lattice~\cite{mks05}.  Unlike the $1+1$ dimensional case, both the
strands of the DNA can get adsorbed on the surface and still satisfy the
non-crossing constraint on the plane ($y$-direction).

The two energies independently give us two special temperatures:
{\it(i)} $T_{\mathrm w}$ the temperature for desorption of the DNA from
the surface, and {\it(ii)} $T_{\mathrm m}$ the melting temperature of
dsDNA.  In absence of a surface, the melting temperature is given by
$k_B T_{\rm m}=\epsilon_b/\ln(4/3)$~\cite{maren,KapriSMB,scaling}. For
ssDNA, $k_B T_{\mathrm w}= \epsilon_w/\ln 2$~\cite{kapri:pre05}. Thermal
fluctuations create bubbles in the dsDNA changing its effective elastic
behaviour with concomitant rise in the desorption temperature.  We
consider energies such that $T_{\mathrm w}<T_{\mathrm m}$ and for
numerical results we choose units so that $k_B=1$,  and $\epsilon_b=1$.
This convention is also adopted in the following discussion, unless we
want to show a general formula or show the dependence on $\epsilon_b$.

There are four distinct phases differentiated by $\langle x_{i}\rangle$
for $i=1,2$ as $N \rightarrow \infty$, ($\langle ...\rangle$ denotes
thermal averaging).  {\it(i)} \phI: zipped DNA adsorbed on the surface
with $\langle x_{i} \rangle/N \rightarrow 0$ for $i=1,2$.  {\it(ii)}
\phII: zipped DNA desorbed from the surface with $\langle x_{i}
\rangle/N =O(1)$, for $i=1,2$.  {\it (iii)} \phIII: unzipped DNA with
the free strand adsorbed on the surface and the pulled strand is
stretched in the direction of the force. This phase is characterized by
$\langle x_1\rangle/N \rightarrow 0$ but $\langle x_2 \rangle /N =O(1)$.
{\it (iv)} \phIV: unzipped DNA with both the strands desorbed from the
surface with $\langle x_2 \rangle /N= O(1)$ and $\langle x_1
\rangle/\sqrt{N}= O(1)$.  The adsorbed or the zipped phases may also be
characterized by the fraction of monomers in contact, e.g.,
$\Phi_{\mathrm w}$ the fraction of polymers in contact with the wall and
$\Phi_{\mathrm b}$ the fraction of bound base pairs.  In a zipped phase
$\Phi_{\mathrm b}\neq 0$ while for the adsorbed phase $\Phi_{\mathrm
w}\neq 0$\footnote{There is a fifth phase U$_{\rm aa}$ in $2+1$
dimensions where DNA melts but both strands remain adsorbed on the
surface. See, e.g.,~\cite{Meunier2003} for experimental determination of
such melting temperatures of adsorbed oligomers on gold.}.

One can do a zero temperature ($T=0$) analysis of the problem, keeping
$\epsilon_b$ constant. The energies of the three phases, namely \phI,
\phII, and \phIII ~are respectively given by $E_{{\mathrm Z_a}}
=-N(\epsilon_w/2 + \epsilon_b)$, $E_{\mathrm{Z_d}} = -N(g +\epsilon_b)$,
and $E_{{\mathrm U}_{\mathrm ad}} = -N( \epsilon_w/2 + g)$.  For
$\epsilon_b <\epsilon_w < 2\epsilon_b$, the phase \phIII ~is always
unstable (i.e. it has higher energy). The transition from phase \phI ~to
phase \phII ~occurs at $g =\epsilon_w / 2$.  On the other hand, for
$\epsilon_w = 2\epsilon_b$, there is a degeneracy for \phII ~and \phIII
~which occurs at $g = \epsilon_w/2$.  But, for $\epsilon_w >
2\epsilon_b$, \phII ~phase is no longer favourable.  The change in
stability of phases \phII ~and \phIII ~as $\epsilon_w$ is tuned is an
indication that these phases could be stabilized by entropy at
intermediate temperatures under appropriate conditions.  For
$\epsilon_w=0$, we can only have \phII ~and \phIV, while, for
$\epsilon_w>0$, as the $T=0$ analysis shows, phase \phI ~must exist. For
very high temperatures $T \gg T_{\mathrm w}, T_{\mathrm m}$, the stable
phase is \phIV.  When $\epsilon_w = \infty$, the free strand, which
remains adsorbed on the surface at all temperatures, acts like a zig-zag
hard-wall. In this case, only phases \phI ~and \phIII ~survive in the
phase diagram.  Consequently, there is a continuous evolution of the
phase diagram for DNA as $\epsilon_w$ is changed.  These phases can
therefore be represented in a 3-dimensional $g$-$T$-$\epsilon_w$ phase
diagram.  We show the cross-sections ($g$-$T$ plane) of this phase
diagram for various $\epsilon_w$ in Fig.~\ref{fig:phdia}.

\figphdia

We like to point out the generic nature of the results, especially the
existence of the phases and the nature of the phase transitions.
Similarly, the choice of a straight wall is not a restriction; a zig-zag
wall with all monomers getting absorbed also show similar
behaviour\cite{Kapri}.  For an experimental realization, DNA tethered to
a gold or mica surface looks promising.  For example, mica-DNA
interaction can be fine tuned by NaCl and MgCl$_2$ over a broad range of
0.02ev per bp to 0.35ev per bp\cite{mica}. For gold 1 kbp DNA adsorption
has also been studied\cite{sensor}.  Unzipping force force measurements
via atomic force microscopy for such finite chains with proper finite
size analysis\cite{KapriSMB:JPCM} could verify the theoretical results
presented here.

Let $D_{n}(x_1,x_2)$ be the partition function (temperature dependence
not shown explicitly) of a dsDNA in the fixed distance ensemble where
$n$th monomers of the strands are at positions $x_1$ (free strand) and
$x_2 \ (x_2 \ge x_1)$ (pulled strand) respectively from the wall.
$D_{n}(x_1,x_2)$ satisfies the recursion relation ( $x_2 \ge x_1 \ge 0$)
\begin{subequations}
\begin{eqnarray}\label{recpfn}
        D_{n+1}(x_1, x_2) &=& \sum_{i,j=\pm 1} D_{n}(x_1 + i, x_2 + j)
        \nonumber \\
        & & \times \left[1 + {\mathcal W}\delta_{x_1,0}\right] \left[1
        + {\mathcal B}\delta_{x_1,x_2}\right],
\end{eqnarray}
where,
\begin{equation}
{\mathcal W} = (e^{\beta \epsilon_w} - 1), {\mathcal B} =
(e^{\beta \epsilon_b} - 1), \ {\rm and}\ \beta = 1/k_BT.
\end{equation}
\end{subequations}
The initial
condition $D_{0}(x_1, x_2) = \delta_{x_1,0}\delta_{x_2,0}$.  The
canonical partition function with an external force $g$ at the end of
the pulled strand is then obtained by summing over all the allowed
configurations of the DNA of length $N$ on the lattice.
\begin{equation}\label{recpfnff}
        Z_N(\beta, g) = \sum_{x_2 \ge x_1 \ge 0}  D_N(x_1, x_2) \
        e^{\beta g x_2}.
\end{equation}
From the partition function we calculate the endpoint averages
$\langle x_1\rangle$ and $\langle x_2\rangle$. The appropriate
response function is the isothermal extensibility, which can be
expressed in terms of fluctuations of the position of the end monomer
\begin{equation}\label{eq:ext}
        \chi = \left. \frac{\partial \langle x_{2} \rangle
        }{\partial g} \right |_{T} =  \frac{1}{k_B T} \left [ \langle
        x_{2}^{2} \rangle - {\langle x_{2} \rangle}^2 \right ].
\end{equation}
The $N$ (length) dependence and finite size scaling of these
quantities would be utilized to identify the phases and the phase
transitions with the help of the Bhattacharjee-Seno data collapse
program\cite{bh:seno}.  (See Fig.~\ref{fig:tu}(a) as an example of the
data collapse of $\chi(g,T)$.)  The nature of the transition (first or
second order) is inferred from the values of the relevant
exponents\cite{KapriSMB:JPCM}.  The phase diagrams are obtained by the
repeated use of finite size scaling.

\figtu

In Fig.~\ref{fig:xg}(a), we have shown the force-distance isotherms for
$\epsilon_w = 1.8$ at two different temperatures $T = 0.5$ and $1.5$ for
the chain of length $N=1000$. In both cases, we start from the ground
state at $g=0$, an adsorbed DNA on the surface. The phases can be
identified by the extensivity of $\langle x_1\rangle$ and/or $\langle
x_2 \rangle$.  At $T = 0.5$, there is a critical force, $g_{\rm s}$, at
which the DNA gets unzipped from the surface but remains double
stranded. We call this as ``transition Sz''.  But at $T=1.5$, we see the
sequence \mbox{ \phI ~$\Longleftrightarrow$ \phII ~$\Longleftrightarrow$
\phIII }, an additional transition (to be called ``transition Uz'') at
$g=g_u$.  The isotherms are obtained at $T=1.5$ by two different methods
with results comparing nicely.  The lines are from the exact transfer
matrix based on Eq.  (\ref{recpfn}), whereas the bigger symbols (squares
and circles) are obtained by performing Monte Carlo simulations for
longer chains using the multiple histogram technique~\cite{ferren89}.
The details of Monte Carlo simulation will be discussed elsewhere.  The
estimates, so obtained, are shown by the upper and the lower triangles
for the free and the pulled strand respectively in Fig.~\ref{fig:xg}(a).
An isotherm for a $2+1$ dimensional case is also shown in the Fig.
~\ref{fig:xg}(b) and is similar to the 2-dimensional case, as already
mentioned.

Figure~\ref{fig:phdia}(a) shows the phase diagram for $\epsilon_w=1$ as
a representative in $0 < \epsilon_w \le \epsilon_b=1$. It contains three
phases, namely \phI, \phII ~and \phIV.  The phase boundary separating
phase \phI ~from phase \phII ~is shown by circles.  The minimum
temperature, $T_{\rm u}$, above which the unzipping of the dsDNA to two
single strands takes place is at $T_{\rm u} = \epsilon_b/\ln 2$ same as
for $\epsilon_w = 0$ case. For low values of $g$, the pulled strand is
not necessarily straight.  If we ignore the effect of the wall the phase
boundary can be calculated exactly from the above recursion relation as
\begin{equation}
	\label{eq:3}
	g_u(T)= \frac{k_B T}{2} \ \ln \left( \frac{ 2e^{-\beta\epsilon_b} -
	2} { 1 - 2e^{-\beta\epsilon_b}} \right).
\end{equation}
Below $T_{\mathrm u}$, the DNA remains double stranded for any value
of force $g$.  That the effect of the wall is negligible is borne out
by the excellent agreement~\cite{Kapri} of the phase boundary for
$\epsilon_w\le 1$ (Fig. \ref{fig:phdia}a).

\figxg

The situation of interest is $1 < \epsilon_w < 2 $.  In this case, we
have all the four phases in the phase diagram, including the $T=0$
unstable phase \phIII.  With $\epsilon_w > \epsilon_b=1$, pair breaking,
as opposed to desorption, would play an important role at low
temperatures. As a result, the phase boundary separating phase \phI
~from phase \phII, loses its monotonicity seen in the $\epsilon_w = 1$
case and a thin slice of reentrance starts appearing in the phase
diagram at intermediate temperatures.  Such an intermediate reentrance
was also found in Ref.~\cite{KapriSMB}, in a different context of a
dsDNA with a force at an interior point. We have shown the phase diagram
for $\epsilon_w = 1.8$ in Fig.~\ref{fig:phdia}(b), as a representative
of this regime with the reentrance shown in the inset.  Apart from this
feature, there is a region in the phase diagram which involves three
phases, namely \phI, \phII ~and \phIII.  The transitions from phase
\phII ~to phases \phIII, and \phIV ~are of first order, whereas, the
transition from phase \phIII ~to phase \phIV ~is second order.  The
critical line is at $T_{\mathrm e} = \epsilon_w / \ln 2$ for all $g$.
This is the temperature at which the adsorbed free strand in phase
\phIII ~desorbs from the surface.  If we ignore the effect of attraction
of the wall on phase \phII, the first order phase boundary can be
obtained by equating the free energies of the two phases obtained from
Eq.~(\ref{recpfn}).  For the \phII ~to \phIII ~transition we then get
\begin{equation} 
	\label{eq:4} 
	g_{\ell}(T)=\frac{k_B T}{2}\ \ln \left [ \frac{ \left(
	e^{-\beta\epsilon_w} - e^{-2\beta\epsilon_w} \right)^{1/2} }
	{e^{\beta\epsilon_b} - \left( e^{-\beta\epsilon_w} -
	e^{-2\beta\epsilon_w} \right)^{1/2} } -1   \right ],
\end{equation}
while the \phII ~to \phIV ~transition is still given by
Eq.~(\ref{eq:3}). Of course Eq.~(\ref{eq:4}) will not be appropriate
in the temperature region where there is  reentrance in the desorption
boundary.  This is seen when compared  with the numerical results
(Fig.~\ref{fig:phdia}b).
However the large force asymptote is given correctly, especially the
dependence of $T_{\mathrm u}$ on $\epsilon_w$ as
\begin{equation}
	\label{eq:5}
	\epsilon_w = k_B T_{\mathrm u} \ \ln \left[ \frac{2} { 1 - \sqrt{ 1
	- 4 \exp \left(-\epsilon_b / k_{\mathrm B}T_{\mathrm u} \right) }}
	\right].
\end{equation}
This expression has the correct limits and matches with the numerics
as shown in Fig.~\ref{fig:tu}(b), justifying {\it a posteriori} the neglect
of the effect of the attractive wall in these conditions.

The free strand desorption critical line terminates on the first order
boundary for phase \phII ~separating it from phases \phIII ~and \phIV.
This point of intersection is the CEP. Note that $g_{\ell}(T)$ and
$g_u(T)$ meet at $T_{\mathrm e} =\epsilon_w/\ln 2$ with same slope as
they should for a CEP.  This point is shown in Fig.~\ref{fig:phdia}(b),
by C$_{\mathrm e}$ (filled square).  For $\epsilon_w=1.8$, C$_e$ is at
$g_{e}=0.742618..., T_{e}=2.59685...$ (with $\epsilon_b=1$). The CEP
appears in the phase diagram only for $\epsilon_w > 1$, and shifts
towards the melting point as $\epsilon_w$ is increased.  That the two
curves meet at CEP with the same slope is a confirmation of its nature.
At a triple point the angle between any two phase boundaries is strictly
less than $2\pi$.

The phase diagram for $\epsilon_w = 2$ has a new feature that \phII
~just becomes unstable at $T=0$.  For $\epsilon_w>2$, a triple point
appears in the phase diagram where \phI, \phII ~and \phIII ~coexist.
The CEP still persists but \phII ~is now stabilized by entropy.  With
the increase of $\epsilon_w$ beyond 2, the region representing phase
\phII ~shrinks rapidly and both the triple point and the CEP shift
towards higher temperatures and disappear independently from the phase
diagram. The crossing of the thermal desorption and thermal melting
temperature introduce new complications.  These will be discussed
elsewhere.

In Fig.~\ref{fig:xg}(c), we have plotted the scaled distances $\langle
x_{1,2} \rangle / N$, of the end monomers of both the strands from the
surface as a function of temperature $T$ at an external applied force
$g=0.925$.  This particular value of the force lies in a small region
which allows us to see all the possible phases, including the
reentrance between the phase \phI ~and the phase \phII ~(see
Fig.~\ref{fig:phdia}(b)). Just by increasing the temperature, the DNA
can be made to go through the sequence 
$$ {\mbox{\phII}}
\Longleftrightarrow {\mbox{\phI}} \Longleftrightarrow {\mbox{\phII}}
\Longleftrightarrow {\mbox{\phIII}} \Longleftrightarrow
{\mbox{\phIV}}.$$ 
In the last phase, the free strand of the DNA desorbs
from the surface and stays at a distance of $\sqrt{N}$ from the
surface (not visible in this scale). To make it visible, we have
plotted, in the inset, the scaled separation, $\langle x_1 \rangle /
\sqrt{N}$, of the end monomer of free strand from the surface as a
function of $T$ for DNA of various lengths.  The plot confirms the
existence of such a transition.

The adsorption-desorption transition in the model is a classical
second-order transition.  One therefore expects a Landau type theory
to be applicable for the CEP.  A CEP is described by a eighth-order
Landau function $ F = t \phi^2 + \phi^4 + w\phi^6 + \phi^8$ in terms
of a suitable order parameter~\cite{helena}.  The first order transition
takes place between two ordered phases ($\phi\neq 0$) with $w>0, t<0$
or between an ordered phase ($\phi\neq 0$) and a disordered phase for
$w<0,t>0$. A second order transition takes place between the other
ordered phase and the disordered phase at $t=0$. The CEP is at $t=0,
w=-3/\sqrt{2}$.  The first order line develops a singularity
associated with the behaviour of the specific heat across the critical
line. Even though the first order lines are continuous with the same
slope, the curvatures are different because the specific heat has a
jump discontinuity across the critical line.

By fitting the first order boundaries near the CEP at $(T_{\mathrm e},
g_{\mathrm e})$ (for $\epsilon_w=1.8$), we find
\begin{equation}
	\label{eq:2}
	g_{\ell , u}(T)= a (T-T_{\mathrm e}) + b_{ \ell, u} (T-T_{\mathrm
	e})^2 + \dots,
\end{equation}
with $a=-0.87 \pm 0.01$, $b_{\ell}=-0.49 \pm 0.03$, $b_u=0.024 \pm
0.019$, where the subscript $u$ is for $T>T_{\mathrm e}$ and $\ell$ for
$T<T_{\mathrm e}$. This jump in the second-order is consistent with the
prediction of the Landau theory. Fig. \ref{fig:cecurv} shows the fits
near the critical end point.

\figcecurv

To summarize, we established the possibility of four different phases in
a set up to unzip an adsorbed dsDNA by pulling a single strand.  We find
that depending upon the relative strengths of the binding on the surface
$\epsilon_w$ and the pairing energy $\epsilon_b$, either all the four
phases or a few of them, are present in the phase diagram. By keeping
$\epsilon_b$ fixed, we find that a critical end point is present in the
phase diagram for a wide range of $\epsilon_w$. Furthermore, for a
narrow range of $\epsilon_w$, we also have a triple point in the phase
diagram.  It seems that the unzipping of an adsorbed dsDNA by pulling a
single strand can be a potential candidate to explore the critical end
point and this will open up a new vista for single molecular
spectroscopy.

{\it Note Added in Proofs}: In a recent paper Marenduzzo {\it et
al.}~\cite{maren2008} also obtained eq.~(\ref{eq:4}) in a different
context of melting of a stretched DNA.

\end{document}